\documentstyle[preprint,aps,prl, amssymb,amsmath]{revtex}

\begin{document}
\draft

\title{Negative Domain Wall Resistance in Ferromagnets}

\author{R.~P.~van~Gorkom\cite{RvG}, Arne~Brataas\cite{philips}, and
Gerrit~E.~W.~Bauer}

\address{Department of Applied Physics and Delft Institute of
Microelectronics and Submicrontechnology, Delft University of
Technology, Lorentzweg 1, 2628 CJ Delft, The Netherlands}

\date{\today} \maketitle

\begin{abstract}
The electrical resistance of a diffusive ferromagnet with magnetic
domain walls is studied theoretically, taking into account the spatial
dependence of the magnetization.  The semiclassical domain wall
resistance is found to be either negative or positive depending on the
difference between the spin-dependent scattering life-times.  The
predictions can be tested experimentally by transport studies in doped
ferromagnets.
\end{abstract}


A domain wall (DW) is the region between two ferromagnetic domains in
which the direction of the magnetization rotates.  A number of
experiments have been conducted which show either an {\em increase}
\cite {Gregg,Viret,Ruediger99} or a {\em decrease}
\cite{Hong,Otani,Ruediger98,Theeuwen,Taniyama} of the resistance due
to DWs compared to the resistance of a single domain ferromagnet.
These experiments have been done on thin films, structured thin films,
and membranes, in the diffusive transport regime where the electron
mean free path is shorter than the typical system size.

In the diffusive limit, Cabrera and Falicov\cite{Cabrera74:217}
calculated an {\em increase} of the resistance caused by the
back-reflection of electrons by the domain wall.  The reflection
probability was found to be exponentially small in the ratio of the DW
width to the Fermi wavelength.  An {\em increase} of the semiclassical
resistance has also been predicted by Tatara and Fukuyama
\cite{Tatara} by linear response calculations assuming
spin-independent relaxation times.  Levy and Zhang\cite {Levy}
obtained the DW resistance from a Boltzmann equation.  They showed
that spin-dependent relaxation times can enhance this {\em positive}
DW resistance, depending on the ratio of relaxation times of the
majority and minority spin electrons.  Brataas et al.  \cite{Brataas}
calculated the domain wall resistance generalizing the approach of
Tatara and Fukuyama to include spin-dependent lifetimes with
qualitively similar results to Levy and Zhang.

The only intrinsic mechanism which explains a {\em decrease} of the
resistance has been proposed by Tatara and Fukuyama \cite{Tatara},
viz.\ the destruction of electron weak localization by the dephasing
caused by the domain wall decreases the resistance.  However,
experimentally, the negative domain wall resistance persists up to
relatively high temperatures \cite{Otani,Ruediger98} where
localization does not play a role.  Kent et al.\cite{Ruediger98}
explain the negative DW resistance by an extrinsic effect: reduced
surface scattering.

It is the purpose of this Letter to show that the semiclassical DW
resistance of diffusive ferromagnets can be {\em negative} as well as
{\em positive} when the electronic structure of the domain wall is
taken into account semiclassically.  The experimental results \cite
{Gregg,Viret,Ruediger99,Hong,Otani,Ruediger98,Theeuwen,Taniyama} may
thus originate from one and the same intrinsic semiclassical effect.

Let us first describe the elementary physics of electron transport in
a domain wall. The Drude resistivity of a single-domain ferromagnet
reads in the 2-band Stoner model \cite{Boltzman}
\begin{equation}
\rho=\frac{m}{e^2} \frac{1}{n_{+}\tau_{+}+n_{-}\tau_{-}},
\end{equation}
where $m$ is the mass of the electron, $e$ is the charge of an
electron, $n_{+}$ $(n_{-})$ is the density of spin-up (down) electrons
and $\tau_{+}$ $(\tau_{-})$ is the scattering relaxation time for the
spin-up (down) electrons which at low temperatures depends on the
(spin-dependent) impurity potential and (spin-dependent) density of
states.  A redistribution of the electrons between the spin-up and
down bands (i.e.  a change in magnetization) modifies the resistivity
when $\tau_{+}\ne \tau_{-}$.  With $ n = n_{+}+n_{-}$,
$n_{\pm}=n_0^{\pm}+\delta n_{\pm}$ and $\delta n_{+}=-\delta n_{-}$,
the change in resistivity is found to be:
\begin{align}
\delta \rho \approx - \rho_0^2 \frac{e^2}{m} \delta n_{+} (\tau_{+} -
\tau_{-}) \, ,
\label{simpleresis}
\end{align}
where $\rho_0$ is the resistivity of a single domain ferromagnet.  We
see that a modified magnetization causes the resistivity to either
increase or decrease, depending on the relaxation times.  The
relaxation times in a ferromagnet depend on the type of impurities
that are present in the material \cite{Campbell}.  The sign and the
magnitude of the resistivity change are therefore impurity specific.
In the following we show that the magnetization is modified in a
domain wall and contributes to the DW resistance on top of the DW
scattering mechanisms discussed in literature
\cite{Cabrera74:217,Tatara,Levy,Brataas}.

We approximate the domain wall by a local constant rotation along the
$z$-direction\cite{Levy}
$\partial_{z}\phi (z)=\pi/\lambda_w=a_0$,
where $\lambda_w$ is the length of the domain wall, as indicated in
Fig.\ \ref{DWdrawing}.  This is allowed when the domain wall is much
wider than the Fermi wave length.  The DW resistance can in this limit
be calculated by interpreting the DW as a finite slice of a so-called
spin-spiral ferromagnet.  The total resistance of a ferromagnet is
determined by simply adding the resistivities of the DWs and the
domains.  The relative change in resistance due to the domain walls is
$(R-R_0)/R_0=(L\rho_0 + \lambda_w (\rho_0+\delta \rho)
-(L+\lambda_w)\rho_0)/((L+\lambda_w)\rho_0)=\lambda_w \delta
\rho/((L+\lambda_w) \rho_0)$,
where $L$ is the length of the domain.  

The magnetization is a result of the exchange interaction.  We follow
the common procedure to express the exchange energy by the continuum
limit of the mean-field Heisenberg model
\begin{equation}
E_{\text{ex}}= \int d{\bf r} \int d{\bf r}' K(|{\bf r}-{\bf r}'|)
\langle {\bf m}({\bf r}') \rangle \cdot {\bf m} ({\bf r}) \, ,
\label{exchange}
\end{equation}
where $K(|{\bf r}-{\bf r}'|)$ is the exchange interaction between
electrons in a volume element ${\bf r}$ and ${\bf r}'$, ${\bf m}({\bf
r})=\mu_B {\bf s}({\bf r})$ is the magnetization operator and $\langle
{\bf m}({\bf r}) \rangle$ denotes the thermal ensemble average of the
magnetization. The range $\lambda_{\text{ex}}$ of the exchange
interaction is of the order of the Fermi wavelength. In a DW
neighboring spins are canted.  Hence the exchange energy is
reduced. Using $\langle {\bf m}({\bf r}') \rangle \cdot {\bf m}({\bf
r})=\cos(\pi (z-z')/\lambda_w) \langle {\bf m}({\bf r}) \rangle \cdot
{\bf m}({\bf r})$ and taking into account that $\lambda_w \gg
\lambda_{\text{ex}}$, the exchange energy becomes $E_{\text{ex}} =
\int d{\bf r} {\bf H}_{\text{ex}}({\bf r}) \cdot {\bf m}({\bf r})$,
where the exchange field is
\begin{equation}
{\bf H}_{\text{ex}}({\bf r}) \approx K \left[ 1-\frac{1}{2} (\frac{\pi
\lambda_{\text{ex}}}{\lambda_w})^2 \right] \langle {\bf m}({\bf r})
\rangle \, ,
\label{splitting}
\end{equation}
and $K$ is the total exchange integral.  The exchange field and therefore 
the splitting decrease with decreasing $\lambda_w$ and increasing 
$\lambda_{\text{ex}}$.  The effect of the reduced magnetization on the 
resistivity seems small, since the domain wall is much wider than the Fermi 
wavelength.  However, we will show that the effect of the reduced 
magnetization on the DW resistance is of the same order as that of other 
mechanisms studied previously\cite{Tatara,Levy,Brataas}.  Expressing the 
exchange splitting as $2\Delta({\bf r})=2J s({\bf r}) $, we have $J=\mu_B 
K(1-(\pi \lambda_{\text{ex}}/\lambda_w)^2/2)$ (the thermal average 
spin-density is s({\bf r})=$|\langle {\bf s}({\bf r})\rangle|$).  The 
relative change in the effective coupling constant $J$ due to the domain 
wall can be written as $\delta J/J=-\kappa 2 E_w / 
(\epsilon_F^{+}+\epsilon_F^{-}))$, where $E_w=\hbar^2 a_0^2/(2m)$ is an 
energy parameter of the rotation of the magnetization, and 
$\epsilon_F^{\pm} = \hbar^2 (k_F^{\pm})^2/(2m)$ is related to the (spin 
dependent) Fermi wavevectors.  Using an estimate for 
$\lambda_{\text{ex}}\approx \lambda_F/2 = \pi /k_F$ we find 
$\kappa=\pi^2/2$.  A longer range of the exchange interaction 
(\ref{exchange}) will have even larger effects.  Note that $\Delta$ 
decreases even faster than $\delta J/J$ because in the self-consistent mean 
field approximation $|s|$ is also reduced in the DW, as will be shown 
below.  Other effects that may change the magnetization and therefore also 
the resistivity are magnetostriction and the internal dipolar magnetic 
field.  The former can change the exchange integral due to a change in 
lattice constant, caused by the domain wall.  The latter directly affects 
the splitting.  However we expect these effects to be smaller than those 
discussed here.

The electronic structure of a two band ferromagnet with non-collinear
magnetization can be found from the Stoner Hamiltonian
\begin{equation}
H= -\frac{\hbar ^{2}}{2m}
\nabla ^{2}+\mu_{B}{\bf H}_{\text{ex}}({\bf r})\cdot \text{\boldmath
$\sigma$}, \label{ham}
\end{equation}
where the three componets of $\text{\boldmath$\sigma$}$ are the Pauli
spin matrices ($\sigma _{x},\sigma _{y}$ and $\sigma _{z}$) and ${\bf
H}_{\text{ex}}({\bf r})$ is the exchange field as described above.  We
disregard the spin-orbit interaction and the Lorentz force due to the
internal magnetization, because the DW magnetoresistance can
experimentally be separated from the anisotropic magnetoresistance
(AMR) and the ordinary magnetoresistance (OMR) \cite{Ruediger98}.

We solve the eigenvalue problem for the Hamiltonian Eq.\ (\ref{ham})
 by a local gauge transformation, assuming translational symmetry in
 the $x$ and $y$ directions and introducing the Fourier transform of
 the gradient of the magnetization direction
$\partial _{z}\phi (z)=\sum_{q}\exp (iqz)a_q$
\cite{Tatara,Brataas}, where $\phi (z) $ is the angle of magnetization
 in the rotation plane. After the gauge transformation the Hamiltonian
 (\ref{ham}) becomes $\widetilde{H}=H_{0}+V$, where $H_{0}=\sum_{{\bf
 k}\sigma }\left( \epsilon _{{\bf k}\sigma }-\mu \right) c_{{\bf
 k}\sigma }^{\dagger }c_{{\bf k}\sigma }$ ($\epsilon _{{\bf k} \sigma
 }=\hbar ^{2}k^{2}/2m-\sigma \Delta $) and the interaction $V$ with
 the DW is specified in Ref.\ \cite{Tatara,Brataas}.

We proceed by calculating
$\delta \rho$, the difference in resistivity of the spin spiral
compared to the single domain state.  In the spin-spiral $\partial
_{z}\phi$ is constant everywhere.  The Hamiltonian (\ref{ham}) is
diagonalized in spin space by ${\bf u} _{\pm }={\cal N}_{\pm}[1,i(1\mp
\sqrt{1+\alpha ^{2}})/\alpha ]^{T}$, where $ {\cal N}_{\pm }$ is a
normalization constant, $\alpha =k_{z}a_{0}/p^{2}$, and $p^2=2m \Delta
/\hbar^2$ \cite{Brataas}.  The eigenvalues are:
\begin{eqnarray}
E_k^{\pm }=\frac{ \hbar ^{2}}{2m} \left( k^{2}+a_{0}^{2}\mp
\sqrt{k_{z}^{2}a_{0}^{2}+p^{4}}\right ) .
\label{energy}
\end{eqnarray}
Due to the rotation of the direction of the spin-quantization axis
\begin{equation}
u_{\pm}^{\dag }\sigma _{z}u_{\pm }=\pm \frac{1}{\sqrt {1+\alpha ^{2}}},
\label{redspind}
\end{equation}
the spin-density in the direction of the local magnetization becomes
\begin{eqnarray}
s_\pm=\pm\frac{1}{V}\sum_{k }\frac{1}{\sqrt{1+\alpha ^{2}}}f(E_k^\pm -\mu ),
\end{eqnarray}
whereas the electron densities of the spin-up and down eigenstates remain as
$n_\pm=(1/V)\sum_{k}f(E_k^\pm-\mu )$.  The total spin-density is $s=s_+ + 
s_-$.

At $T=0 K$ we find
\begin{eqnarray}
n_\pm=\frac{1}{6 \pi^2} (k^\pm_1)^3 \left ( 1 \pm \frac{E_w}{4 \Delta
} \right )
\end{eqnarray}
and
\begin{eqnarray}
s_\pm=\pm \left ( n_\pm - \frac{1}{30 \pi^2} (k^\pm_1)^5 \frac{\hbar^2
E_w}{\Delta^2 4 m} \right ),
\label{spinpolar}
\end{eqnarray}
where $k_1^\pm=( 2m/\hbar^2 ) ^{1/2} \left ( \mu \pm \Delta
-\frac{1}{4}E_w \right ) ^{1/2}$.
Without spin rotation ($a_0 \rightarrow 0$), $n$ and $s$ reduce to the
familiar form $n_0^\pm=\pm s_0^\pm=(1/6 \pi ^2) (2m/\hbar^2) ^{3/2}
(\epsilon_F^\pm)^{3/2}$, where $\epsilon_F^{\pm}=\mu \pm \Delta$.

The numbers of spin-up and spin-down electrons have to be calculated
self-consistently, since the effective spin splitting depends on the
spin densities.  A wide domain wall only weakly modifies the
electronic structure.  Therefore the reduced magnetization can be
calculated by perturbation theory.  Charge neutrality is taken into
account as in Ref.\ \cite{Brataas} introduing a shift in the chemical
potential $\mu= \mu_0 + \delta\mu$. The spin densities are denoted as
$s_\pm=\pm n_0^\pm+\delta s_\pm$, and the splitting of the bands
becomes $\Delta = \Delta _{0}+\delta \Delta $, where $\mu_0$ is the
single-domain chemical potential, $\delta \mu$ is the change in
chemical potential caused by the DW, $\delta s_{+}$ $(\delta s_{-})$
is the change in spin density in the spin-up (down) band, and
$\Delta_0=(n_0^{+}-n_0^{-})J_0$.  The exchange splitting is modified
as
\begin{eqnarray}
\delta \Delta=(|\delta s_{+}|-|\delta s_{-}|)J_0 + (n_0^{+}-n_0^{-})\delta J,
\label{deltaDelta}
\end{eqnarray}
where the first term is due to the reduced spin density in the spin
spiral, and the second term reflects the reduced exchange interaction.
We obtain:
\begin{eqnarray}
\delta n_{\pm}=N_{\pm} \left ( \delta \mu \pm \delta \Delta -
\frac{1}{4}E_w \pm \frac{E_w}{6 \Delta}\epsilon_F^\pm \right )
\label{deltan}
\end{eqnarray}
and
\begin{eqnarray}
\delta s_{\pm}=\pm \left ( \delta n_{\pm} - \frac{E_w}{15
\Delta^2}N_\pm ( \epsilon_F^\pm )^2 \right ),
\label{deltaspinpolar}
\end{eqnarray}
where $N_+$ $(N_-)$ is the density of states of the spin-up (down) band at the
Fermi energy.
With $n_\pm=2 N_\pm \epsilon_F^\pm/3$, which holds for parabolic bands, and
Eqs.\
(\ref{deltaDelta}), (\ref{deltan}), and (\ref{deltaspinpolar}),
we can find the electron densities, which substituted into Eq.\
(\ref{simpleresis}) yield:
\begin{eqnarray}
\delta \rho & = & \frac{e^2\rho_0^2}{m}\frac{E_w}{J_0 D \Delta_0
}(\tau_+-\tau_-) \times
\label{resis}
 \\ &&
 \left (
 \frac{\kappa \Delta_0^2}{\epsilon^{+}_{F} +\epsilon^{-}_{F}}
-\frac{\epsilon^{+}_{F} + \epsilon^{-}_{F}}{24}
+\frac{J_0}{\Delta_0}\frac{  N_{+} (\epsilon^{+}_{F})^2 -
N_{-}(\epsilon^{-}_{F})^2 }{30}
 \right ), \nonumber
\end{eqnarray}
which is our main result.  The first term is directly due to the
reduced exchange interaction, the second term reflects the change in
dispersion of the spin-up and spin-down band (last term in Eq.\
(\ref{energy})), and the third term is due to the reduced spin-density
(Eq.\ (\ref{redspind})).  The first term is almost always bigger than
the other two terms.  Eq.\ (\ref{resis}) shows that the change in
resistivity due to the reduced magnetization is of the same order as
the effect of the spin-flip scattering\cite{Brataas}, both scaling
linearly with $E_w$.  The dimensionless denominator
$D=(N_++N_-)/(4N_+N_-J_0)-1$ appears as a result of the
self-consistency and is always positive (see below).  $D$ vanishes
when the spontaneous magnetization disappears with decreasing $J_0$.
In that case the non-degenerate perturbation is not valid anymore,
because $\delta \rho \to \infty$.  When $\tau _{+}<\tau _{-}$ the DW
resistance is negative!

In order to estimate the importance of the effect, we introduce 
dimensionless variables: $\gamma \equiv k_{+}/k_{-}$ is the ratio of the 
Fermi wavevectors and $\beta \equiv \tau _{+}/\tau _{-}$ is the ratio of 
relaxation times.  The dimensionless denominator then becomes 
$D=(\gamma+\gamma^{-1}-2)/3 > 0 $.  Microscopic theory\cite{Brataas} 
reveals that additional spin flip terms, which always increase the 
resistance, have to be added to the result derived here \cite{note}.  Fig.\ 
\ref{figureresis} shows $\delta \rho/\rho_0$, as calculated from Eq.\ 
(\ref{resis}) (fine dashed line), due to the spin-flip scattering (taken 
from Ref.\ \cite{Brataas}) (coarse dashed line) and the sum of both effects 
(solid line) as a function of the ratio of relaxation times $\beta$, for 
two different ratios of the Fermi wavelengths $\gamma$.  
$E_w/\epsilon_F=4.3 \times 10^{-4}$ for Co has been estimated from 
$\lambda_w=15$ nm and $k_F= 1$ \AA$^{-1} $.  The exchange length is equal 
to $\lambda_{\text{ex}}= 2.8$ \AA.  In Fig.\ \ref{figureresis} we can see 
that the DW resistivity is between -5\% and 10\% for the parameters chosen, 
i.e.  depending on the value of the exchange integral, the impurities, and 
the band structure of the material.  Smaller spin splittings and larger 
asymmetries in the relaxation times increase the domain wall resistance.

Our results agree with the experimental finding that the DW resistance can 
be negative \cite{Hong,Otani,Ruediger98,Theeuwen,Taniyama}.  We have shown 
that the sign of the DW resistance depends on the difference of scattering 
relaxation times, which can be positive or negative.  This difference is to 
a large extent determined by the kind of impurities present in the sample, 
e.g.  theoretical calculations for Cr impurities in Fe give $\beta =0.11$, 
i.e.  $\tau_+ - \tau_-<0$ whereas for Cu impurities $\beta =3.68$, i.e.  
$\tau_+ - \tau_->0$ \cite{Mertig94:11767}.  Also for Ni the DW resistance 
can have both signs, since $ 0.2<\beta <30$ \cite{Mertig94:11767}.  
Experimentally the parameter $\beta$ is not available, but the ratio of the 
resistivities has been determined \cite{Campbell} yielding for Co $0.1< 
\rho_+/\rho_- < 5$, Ni $0.05< \rho_+/\rho_- < 8$, and Fe $0.1< 
\rho_+/\rho_- < 10$, where $\rho_+/\rho_-=\tau_-n_-m_+/(n_+\tau_+ m_-)$.  
(Only $\tau_+$ and $\tau_-$ depend on the type of impurities 
\cite{Stearns}).  Therefore, a reasonable agreement exists between 
experimental and theoretical values of the scattering relaxation times.  
This means that our predictions for the DW resistance can be experimentally
tested by intentionally doping samples with different impurities and 
measuring the DW resistance as a function of type and concentration.

For realistic band structures the ballistic contribution to the domain
wall resistivity from the spin flip terms\cite{Tatara,Levy,Brataas} is
enhanced due to the near degeneracy of the different bands at the
Fermi energy\cite{Brataas98}. Similarly, we expect an enhancement of
the present effect, because the exchange splitting is more sensitive
to the gradient of the magnetization when the bands are nearly
degenerate.

In conclusion we have shown that a negative as well as a positive
domain wall resistance is possible in the semiclassical regime, due to
spin-dependent relaxation times and the spatial dependence of the
magnetization.

This work is part of the research program for the ``Stichting voor
Fundamenteel Onderzoek der Materie'' (FOM), which is financially
supported by the ''Nederlandse Organisatie voor Wetenschappelijk
Onderzoek'' (NWO).  This study was supported by the NEDO joint
research program (NTDP-98).  We acknowledge benefits from the TMR
Research Network on ``Interface Magnetism'' under contract No.
FMRX-CT96-0089 (DG12-MIHT).  We also acknowledge stimulating
discussions with J.~Caro, P.~J.~Kelly, A.~D.~Kent, G.~Tatara,
S.~J.~C.~H.~Theeuwen, and K.~P.~Wellock.

\begin{figure}
\caption{A schematic picture of the magnetization angle as a function
of position.  The dashed line sketches a typical magnetization and the
solid line the piece-wise spin-spiral approximation.}
\label{DWdrawing}
\end{figure}

\begin{figure}
\caption{The relative change in resistivity due to the reduced
magnetization calculated here, the DW spin flip scattering
[\ref{l:Brataas}], and the sum of both effects $[\ref{l:note}]$ as a
function of the ratio of relaxation times. (a) is for the ratio of
Fermi wave vectors $\gamma=1.1$ and (b) $\gamma=2$.  }
\label{figureresis}
\end{figure}


\begin{references}
\bibitem{note}
The calculated resistivity includes for $\delta \rho$ in Ref.\ \cite{Brataas}
already some of the screening effects considered
here and cannot be simply added to the present result.
This is corrected by adding
$\rho_0^2 e^2 (\tau_+ - \tau_-)(\epsilon_F^{+}+\epsilon_F^{-}) E_w (N_{+} 
N_{-})/((N_{+}+N_{-})6m\Delta_0)$
to the sum of the two results.
\label{l:note}
\end{references}
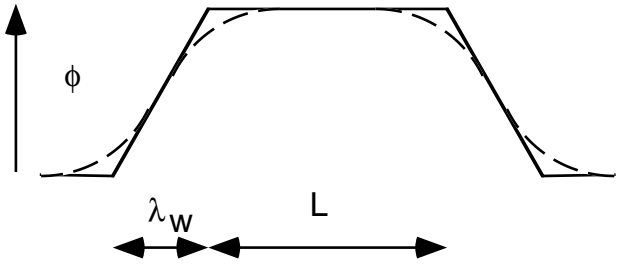
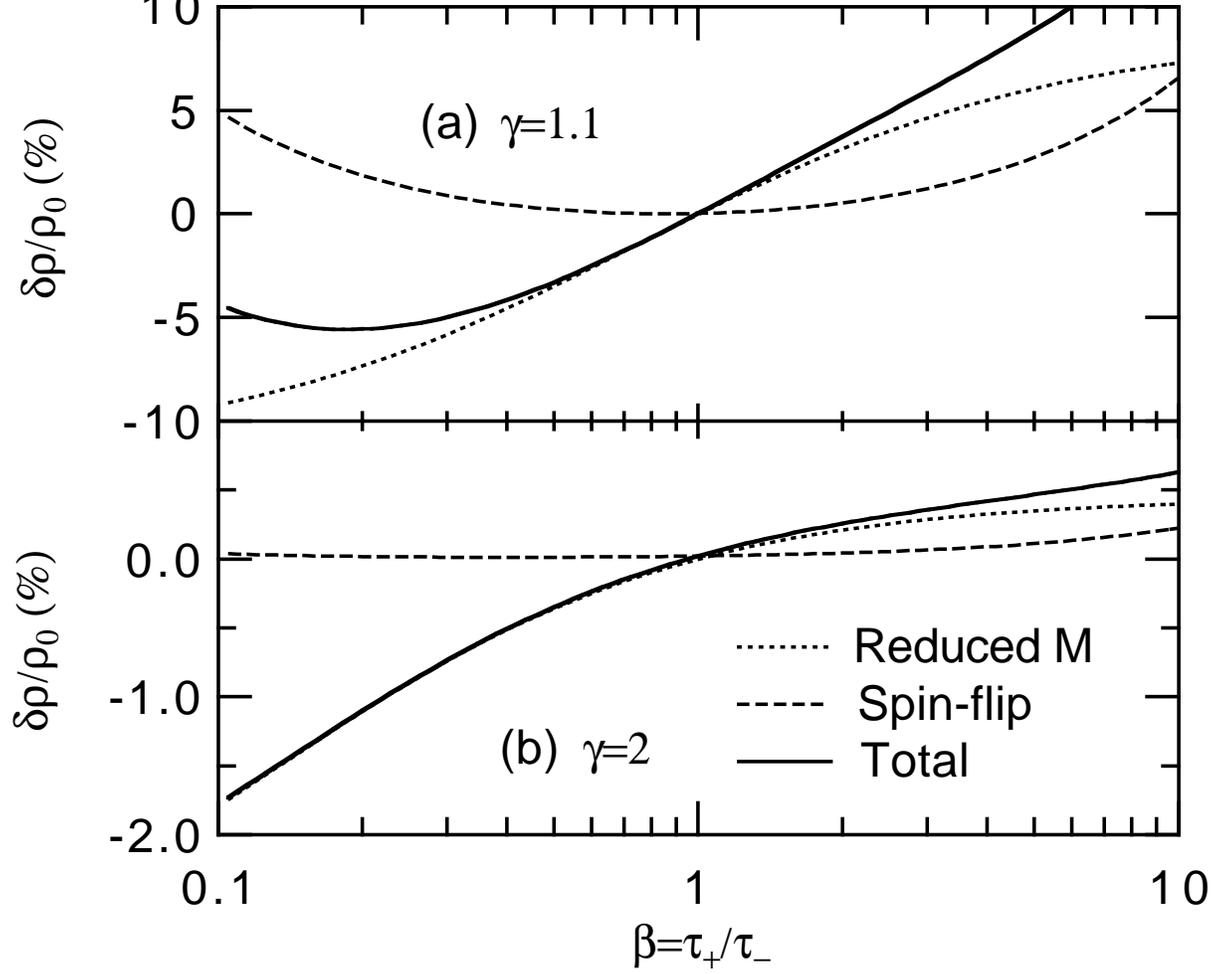
\end{document}